\documentclass[12pt,thmsa]{article}
\usepackage{sw20lart}

\input{tcilatex}

\begin{document}

\title{An $SU\left( 3\right) $ symmetry for light neutrinos}
\author{Riazuddin\thanks{%
email: riazuddin@ncp.edu.pk} \\
National Centre for Physics, Quaid-i-Azam University,\\
Islamabad, Pakistan}
\maketitle

\begin{abstract}
It is proposed that light neutrinos form a triplet in a global $SU(3)$
symmetry in mass eigenstate basis. Assuming that the $SU(3)$ symmetry is
broken in the direction$(-a\lambda _{3}+\frac{b}{\sqrt{3}}\lambda _{8}),$
and after going to the flavor basis we predict atmospheric mixing angle$,$ $%
\sin ^{2}\theta _{23}=0.5$ and $\sin \theta _{13}=0$ if $\nu _{\mu }-\nu
_{\tau }$ symmetry is assumed. In the flavor basis the diagonal part of
matrix coefficient of $b$ (dominant part) is found to transform like $\left(
\lambda _{3}+\frac{1}{\sqrt{3}}\lambda _{8}\right) $. Imposing the same on
matrix coefficient of $a$ fixes solar mixing angle,$\sin ^{2}\theta _{12}=%
\frac{1}{3}$. Implications for neutrinoless double beta-decay are discussed.
\end{abstract}

There is a compelling evidence\cite{r1} that neutrinos change flavor, have
non-zero masses and the neutrino mass eigenstates are different from weak
eigenstates. As such they undergo oscillations.

All neutrino data\cite{r1} with the exception of LSND anomaly\cite{r2} is
explained by three neutrino flavor oscillations with mass squared
differences and mixing angles having the following values\cite{r3}

\[
\Delta m_{solar}^{2}=\Delta m_{12}^{2}=\left( 8.1\pm 1.0\right) \times
10^{-5}\text{eV}^{2} 
\]
\[
\sin ^{2}\theta _{12}=0.30\pm 0.08 
\]
\begin{eqnarray*}
\Delta m_{atm}^{2} &=&\left| \Delta m_{13}^{2}\right| \simeq \left| \Delta
m_{23}^{2}\right| \\
&=&\left( 2.2\pm 1.1\right) \times 10^{-3}\text{eV}^{2}
\end{eqnarray*}
\[
\sin ^{2}\theta _{23}=0.50\pm 0.18 
\]
\begin{equation}
\sin ^{2}\theta _{13}\leq 0.047  \label{p1}
\end{equation}
The above mixing pattern is in conformity with ``bi-tri-maximal'' scheme,
first discussed in \cite{r4}. The neutrino mixing angles are defined by the
lepton mixing matrix\cite{r5}

\begin{equation}
\left( 
\begin{array}{c}
\nu _{e} \\ 
\nu _{\mu } \\ 
\nu _{\tau }
\end{array}
\right) =U\left( 
\begin{array}{c}
\nu _{1} \\ 
\nu _{2} \\ 
\nu _{3}
\end{array}
\right)  \label{p2}
\end{equation}
The matrix $U$ is conveniently parametrized by three mixing angles $\theta
_{12},$ $\theta _{13},$ $\theta _{23}$ and three complex phases the two of
which are so called Majorana phases. We put all phases to be zero.

The purpose of this paper is to study the implications of $SU(3)$ symmetry
for light neutrinos in mass eigenstate basis. $SU(3)$ family symmetry has
previously been used\cite{r6} in a more fundamental way than attempted here.
Our aim is modest. We show that if $SU(3)$ symmetry for neutrino mass
eigenstates is broken in the direction $(-a\lambda _{3}+\frac{b}{\sqrt{3}}%
\lambda _{8})$, and then going to flavor basis by the Unitary transformation
given in Eq. (\ref{p2}), the atmospheric mixing angle is predicted to be: $%
\sin ^{2}\theta _{23}=0.5$ and $\sin \theta _{13}=0$ if $\nu _{\mu }-\nu
_{\tau }$ symmetry\cite{r7} is assumed.This symmetry is inspired by the
experimental observation of near maximal atmospheric mixing angle and small
upper limit on $\theta _{13}$ implying an intresting possibility that there
may be an approximate $\mu \leftrightarrow \tau $ symmetry in the neutrino
sector \cite{r8}. Its deeper origin is not yet known. Such a symmetry has
also intresting implications in Leptogenesis\cite{r9}.

It is found that in flavor basis, the diagonal part of matrix coefficient of 
$b$ (dominant part) exhibits an interesting property, namely, it transforms
like $\left( \lambda _{3}+\frac{1}{\sqrt{3}}\lambda _{8}\right) $. If we
impose the same on matrix coefficient of $a$ (non leading part) also we
predict the solar mixing angle: $\sin ^{2}\theta _{13}=\frac{1}{3}$. In our
approach absolute mass of neutrino in $SU(3)$ limit, $m_{0}$, is not
constrained by neutrino oscillations data. If WMAP constraint on neutrino
mass, $\sum m_{i}<(0.4-0.7)$eV$,)$ is used$,$ the effective double
beta-decay mass $m_{ee}$ $<(0.13-0.23)$ eV.

We assume that mass eigenstates $(\nu _{1},\nu _{2},\nu _{3})$ form an $%
SU(3) $ triplet so that in the $SU(3)$ limit, they have common mass $m_{0}$.
Since the neutrino mass matrix in mass eigenstate basis has to be diagonal
and the only diagonal matrices available are $\lambda _{3}$ and $\lambda
_{8} $, the most general form of symmetry breaking is provided by $%
(-a\lambda _{3}+\frac{b}{\sqrt{3}}\lambda _{8})$ so that in the basis $(\nu
_{1},\nu _{2},\nu _{3}):$%
\begin{equation}
\mathcal{M=}m_{0}I+\left( -a\lambda _{3}+\frac{b}{\sqrt{3}}\lambda
_{8}\right)  \label{I00}
\end{equation}
\begin{eqnarray}
m_{1} &=&m_{0}-a+\frac{b}{3}=m-a  \nonumber \\
m_{2} &=&m_{0}+a+\frac{b}{3}=m+a  \nonumber \\
m_{3} &=&m_{0}-\frac{2b}{3}=m-b  \label{I1}
\end{eqnarray}
where $m=m_{0}+\frac{b}{3}$ and $\left| a\right| $, $\left| b\right|
<<m_{0}. $

Thus 
\begin{eqnarray}
\Delta m_{12}^{2} &\simeq &4ma  \nonumber \\
\left| \Delta m_{23}^{2}\right| &\simeq &2m(a+b)+(b^{2}-a^{2})  \nonumber \\
\left| \Delta m_{13}^{2}\right| &\simeq &2m\left( b-a\right) +(b^{2}+a^{2})
\label{I2}
\end{eqnarray}
The data require that $a\ll b$ so that $m_{3}\ll m_{1}\preceq m_{2}$
(Inverted Heirarchy) and 
\begin{equation}
\left| \Delta m_{23}^{2}\right| \simeq \left| \Delta m_{13}^{2}\right|
\simeq 2mb  \label{I2a}
\end{equation}

We now go to the flavor basis $(\nu _{e},\nu _{\mu },\nu _{\tau })$ by using
Eq. (2) $[s_{1}=\sin \theta _{23},\,c_{1}=\cos \theta _{23},s_{2}\equiv \sin
\theta _{13}]$. The neutrino mass matrix in flavor basis is then 
\[
M_{\nu }=m_{0}I+M 
\]
with 
\begin{eqnarray}
M_{11} &=&\frac{b}{3}\left( 1-3s_{2}^{2}\right) -ac_{2}^{2}\cos 2\theta _{12}
\nonumber \\
M_{22} &=&\frac{b}{3}\left( 1-3s_{1}^{2}c_{2}^{2}\right) -a\left[ \cos
2\theta _{12}\left( -c_{1}^{2}+s_{1}^{2}s_{2}^{2}\right) +\sin 2\theta
_{1}\sin 2\theta _{12}s_{2}\right]  \nonumber \\
M_{33} &=&\frac{b}{3}\left( 1-3c_{1}^{2}c_{2}^{2}\right) -a\left[ \cos
2\theta _{12}\left( -s_{1}^{2}+c_{1}^{2}s_{2}^{2}\right) -\sin 2\theta
_{1}\sin 2\theta _{12}s_{2}\right]  \nonumber \\
M_{12} &=&-bc_{2}s_{2}s_{1}+ac_{2}\left[ c_{1}\sin 2\theta
_{12}+s_{1}s_{2}\cos 2\theta _{12}\right]  \nonumber \\
M_{13} &=&-bc_{2}s_{2}c_{1}-ac_{2}\left[ s_{1}\sin 2\theta
_{12}-c_{1}s_{2}\cos 2\theta _{12}\right]  \nonumber \\
M_{23} &=&-bc_{1}s_{1}c_{2}^{2}-a\left[ s_{1}c_{1}\cos 2\theta _{12}\left(
1+s_{2}^{2}\right) -\sin 2\theta _{12}s_{2}\left( s_{1}^{2}-c_{1}^{2}\right)
\right]  \label{I2b}
\end{eqnarray}

Imposing $\nu _{\mu }\longleftrightarrow \nu _{\tau }$ symmetry, we get 
\[
s_{2}=0\text{, }c_{1}=-s_{1}=\frac{1}{\sqrt{2}} 
\]
Thus $M$ reduces to 
\begin{equation}
\frac{b}{2}\left( 
\begin{array}{lll}
\frac{2}{3} & 0 & 0 \\ 
0 & -\frac{1}{3} & 1 \\ 
0 & 1 & -\frac{1}{3}
\end{array}
\right) -\frac{a}{2}\left( 
\begin{array}{lll}
2\cos 2\theta _{12} & -\sqrt{2}\sin 2\theta _{12} & -\sqrt{2}\sin 2\theta
_{12} \\ 
-\sqrt{2}\sin 2\theta _{12} & -\cos 2\theta _{12} & -\cos 2\theta _{12} \\ 
-\sqrt{2}\sin 2\theta _{12} & -\cos 2\theta _{12} & -\cos 2\theta _{12}
\end{array}
\right)  \label{I3}
\end{equation}

It is intersting to note that the diagonal part of the matrix coefficient of 
$\frac{b}{2}$ transforms as $\lambda _{3}+\frac{1}{\sqrt{3}}\lambda _{8}$,
like electric charge operator of $u$, $d$, $s$ quarks. If we require the
same for the matrix coefficient of $-a/2$, we obtain 
\[
\cos 2\theta _{12}=\frac{1}{3} 
\]
giving 
\[
\sin ^{2}\theta _{12}=\frac{1}{3} 
\]
This is consistent with its experimental value given in Eq.(1). Thus the
neutrino mass matrix in flavor basis is 
\begin{equation}
M_{\nu }=m_{0}I+\frac{b}{2}\left( 
\begin{array}{lll}
2/3 & 0 & 0 \\ 
0 & -1/3 & 1 \\ 
0 & 1 & -1/3
\end{array}
\right) -\frac{a}{2}\left( 
\begin{array}{lll}
2/3 & -4/3 & -4/3 \\ 
-4/3 & -1/3 & -1/3 \\ 
-4/3 & -1/3 & -1/3
\end{array}
\right)  \label{I4}
\end{equation}

The data give 
\[
\frac{a}{b}\simeq \frac{1}{2}\frac{\Delta m_{12}^{2}}{\left| \Delta
m_{23}^{2}\right| }\simeq 1.8\times 10^{-2} 
\]
\begin{equation}
\frac{b}{m_{0}}\simeq \frac{1}{2}\frac{\left| \Delta m_{23}^{2}\right| }{%
m_{0}^{2}}\simeq (1.1)\times 10^{-3}\frac{\text{eV}^{2}}{m_{0}^{2}}
\label{I4a}
\end{equation}
$m_{0}$ is not constrained by oscillation data. However $m_{0}$ is
constrained by WMAP data, $\sum m_{i}<(0.4-0.7)$eV$.$

Thus taking 
\begin{eqnarray}
m_{0} &\simeq &0.1\text{eV},  \nonumber \\
\frac{b}{m_{0}} &\approx &10^{-1}  \label{I4b}
\end{eqnarray}

The $SU(3)$ symmetry thus makes sense as symmetry breaking paramameter is
small.

Finally for neutrinoless double $\beta -$decay, the effective double beta
decay mass $<m_{ee}>$ is predicted to be $[\sigma _{1}$ and $\sigma _{2}$
are Majorana phases, $\theta _{13}=0]$\cite{r7} 
\begin{eqnarray}
m_{ee} &=&\left| \left| m_{1}\right| \cos ^{2}\theta _{12}e^{-2i\sigma
_{1}}+\left| m_{2}\right| \sin ^{2}\theta _{12}e^{-2i\sigma _{2}}\right| 
\nonumber \\
&\simeq &\frac{m_{0}}{3}<m_{ee}<m_{0}  \label{I4c}
\end{eqnarray}

Using WMAP limit on $m_{0}$, we get 
\[
m_{ee}\leq (0.13-0.23)\text{eV} 
\]

In summary a global $SU(3)$ for neutrino mass eigenstates and its breaking
along the direction $(-a\lambda _{3}+\frac{b}{\sqrt{3}}\lambda _{8})$ with $%
a\ll b$ together with $(\nu _{\mu }\longleftrightarrow \nu _{\tau })$
symmetry in the flavor basis and the requirement that the diagonal part of
the neutrino mass matrix $\left( M_{\nu }-m_{0}I\right) $ transforms as $%
\left( \lambda _{3}+\frac{1}{\sqrt{3}}\lambda _{8}\right) $can explain the
data in Eq.(1). Further together with WMAP constraint on neutrino mass,
effective double beta decay mass $m_{ee}$ is predicted.

\textbf{Acknowledgments}

The author acknowledges a research grant provided by the Higher Education
Commission of Pakistan to him as a Distinguished National Professor.

\end{document}